# Normal and anomalous densities in Bose-Einstein condensate with optical lattices

**M. Bendacha and A. Boudjemâa**[1]

*Department of Physics, Faculty of Sciences, Hassiba Benbouali University of Chlef*

*P.O. Box 151, 02000, Chlef, Algeria.*

**Abstract:**

We study the quantum phase transition from the superfluid to the Mott insulator state in two and three dimensional Bose-Einstein condensate (BEC) with optical lattices using Bose-Hubbard Hamiltonian within the Generalized Hatree-Fock-Bogoliubov (GHFB) approximation. The behavior of the depletion and the anomalous fraction has been investigated in the Mott insulator phase. We found that at $T = 0$, these quantities become significant in two and three dimensions. It is shown also that the dimensionality of the lattice enhances the anomalous density.

[1] Corresponding author :e-mail: **a.boudjemaa@univ-chlef.dz**

## 1. Introduction:

Ultracold atoms in optical lattices have proven to be a rich field of investigation both theoretically and experimentally. Periodic potentials are very well known in solid state physics, where the basic description of the system is the Bloch theory for a gas of non-interacting particles, such as electrons in crystalline solids. Cold atoms in optical lattices have opened the possibility to investigate effects not previously observable on ordinary matter crystals, such as Bloch oscillations, Wannier-Stark ladders and Landau Zener tunneling (see for review [1-4]). Since 1999, the research on the ultracold matter was stimulated by the success of the first experiment on Bose-Einstein condensates in optical lattices [5]. Optical lattices are created by the frequency shift due to the interference of laser beams counter-propagating [5], producing periodic potential trap for BEC atoms. In general, optical lattices offer several advantages: first of all, a large number of potentials can be created with almost complete control over the parameters, where they can be altered or switched off entirely during the experiment. Optical lattices are ideally suited to produce and study BEC in one, two and three dimensions. They also permit to adjust the ratio between the kinetic energy and the energy of interaction in the magnetic trap. They lead to study the phase transition from superfluid to a Mott insulator state (see [6, 7]). In general, the creation of optical lattices may provide an opportunity for numerous novel applications and for modeling many effects typical of condensed matter.

On the other side, many experimental and theoretical efforts have been directed towards the behavior of BEC in optical lattices. It has been shown that the behavior of the ultracold atoms can be described by the Bose-Hubbard Hamiltonian when the atoms are cooled to the lowest Bloch band of the periodic potential [8]. The Bose-Hubbard Hamiltonian has been used also to study the quantum phase diagram in optical lattice at zero temperature using two different mean field approximations [9, 10]. The main result of this analysis is that the Bogoliubov approximation fails to predict the transition from superfluid to insulating phase. Moreover, a field theoretical approach in terms of path integral formalism was developed [11] to study the quantum phase transitions and to calculate the second-order quantum corrections to the energy density as well as to the superfluid fraction in cubic optical lattices. Furthermore, numerical calculations have been performed to investigate the ground state and finite temperature phase diagrams by Monte Carlo calculations [12] and by bosonic DMFT method [13].

Our aim in this paper is to investigate the phase transition from superfluid to the Mott insulator in both two and three dimensional optical lattice employing the GHFB approximation which is a suitable ansatz when the optical lattice depth is sufficiently low. We can then assume that the BEC is essentially described by a macroscopic wave function with a small correction terms which represent the noncondensed and the anomalous densities. After that we focus ourselves to study fluctuations in the Mott-insulator phase.

## 2. Formalism

The dynamic of ultracold dilute Bose gas in optical lattice can be described by the Bose Hubbard Hamiltonian which can be written in the grand canonical ensemble as

$$H_{BH} = -t\sum_{\langle i,j\rangle} c_i{}^{+} c_j + \frac{U}{2}\sum_i c_i{}^{+} c_i{}^{+} c_i c_i - \mu \sum_i c_i^{+}\, c_i \ , \tag{1}$$

where the sum in the first term on the right hand side is restricted in nearest neighbors and $c_i{}^{+}$, $c_i$ are respectively the creation and annihilation operators of an atom at the site $i$. The parameter $t$ denotes the hopping at the site $i$ to the nearest neighbor site. $U$ is the strength interaction between two atoms in the same site, where we always assume to be positive in what follows. $\mu$ is the chemical potential. At zero temperature, the Bose-Hubbard model is either in a Mott insulating state for $t \ll U$ , or in a superfluid state when $t \gg U$, or in a supersolid phase where both solid and superfluid orders coexist. Quantum phase transitions in the Bose-Hubbard model were first experimentally observed by Greiner *et al.*[6].

Starting now with the equation of motion for the Bose field operator $c_i$, and its decomposition $c_i = \bar{c}_i + \phi_i$ in term of condensate and non-condensate parts [14]. The condensate wavefunction $\phi_i(\vec{r})$ is defined in Bose systems within the discretized, generalized Gross-Pitaevskii equation

$$i\hbar\dot{\phi_i} = \left[-\frac{\hbar^2}{2m}\Delta + V_{ext}(\vec{r}) - \mu + U(n_0 + 2\tilde{n})\right]\phi_i + U\tilde{m}\phi_i^*, \tag{2}$$

Here $n_0 = |\phi_i(\vec{r})|^2$, $\tilde{n} = \langle \bar{c}_i{}^{+}(\vec{r})\bar{c}_i(\vec{r})\rangle$ and $\tilde{m} = \langle \bar{c}_i(\vec{r})\bar{c}_i(\vec{r})\rangle$ are the condensate, non-condensate and the anomalous densities respectively.

In fact, we may easily show that upon linearizing Eq. (2) around a static solution within the random phase approximation (RPA) [15], we obtain $i\hbar\delta\dot{\phi_i} = \Big[-\frac{\hbar^2}{2m}\Delta + V_{ext}(\vec{r}) - \mu + 2Un\Big]\delta\phi_i + U(\tilde{m} + \phi_i^2)\delta\phi_i^*$. Using the parameterization $\delta\phi_i = \sum_k\left(u_k e^{-i\hbar\omega_k t} - v_k e^{i\hbar\omega_k t}\right)$ in uniform case, we get the coupled Bogoliubov-de Gennes (BdG) equations

$$\begin{pmatrix} \hat{\mathcal{L}} & \widehat{\mathcal{M}} \\ -\widehat{\mathcal{M}} & -\hat{\mathcal{L}} \end{pmatrix} \begin{pmatrix} u_k \\ v_k \end{pmatrix} = \varepsilon_k \begin{pmatrix} u_k \\ v_k \end{pmatrix}, \tag{3}$$

where $\hat{\mathcal{L}} = -\Xi_\mathrm{k} + 2Un - \mu$, $\widehat{\mathcal{M}} = U(\tilde{m} + \phi_i^2)$, $n = n_0 + \tilde{n}$ is the total density and

$\Xi_\mathrm{k} = \left(\mathrm{z}_0 - 2\sum_{i=1}^{d} \cos k_i a\right)t$ with $d$ is the dimension of lattice, $a$ is the lattice parameter and $\mathrm{z}_0$ is the number of nearest neighbors. $u_k$ and $v_k$are the quasiparticle amplitudes which satisfy the constraint $u_k^2 - v_k^2 = 1$, they read:

$$u_k, v_k = \left[\frac{1}{2}\left(\frac{\Xi_\mathrm{k} + Un_0}{\hbar\omega_k} \pm 1\right)\right]^{1/2}, \tag{4}$$

The Bogoliubov energy spectrum is defined as

$$\hbar\omega_k = \sqrt{{\Xi_\mathrm{k}}^2 + 2\Xi_\mathrm{k} U(n_0 + \tilde{m})}\ . \tag{5}$$

Equations (2) and (3) represent a lattice formulation of the finite temperature GHFB formalism.

Moreover, we may include fluctuations in our description of the system inserting the usual Bogoliubov transformation $\bar{c}_i = \sum_k (u_k b_k - v_k b_k^+)$ into the definitions of $\tilde{n}$ and $\tilde{m}$, and using the fact that $2N_k + 1 = \coth\left(\frac{\hbar\omega_k}{2T}\right)$, where $N_k = \langle b_k^+ b_k \rangle = 1/\left(e^{\frac{\hbar\omega_k}{2T}} - 1\right)$ is the Bose-Einstein distribution. Therefore, the noncondensed and the anomalous densities turn out to be given:

$$\tilde{n} = \frac{1}{2}\sum_k \left[\frac{\Xi_\mathrm{k} + U(n_0 + \tilde{m})}{\hbar\omega_k} \coth\left(\frac{\hbar\omega_k}{2T}\right) - 1\right], \tag{6}$$

$$\tilde{m} = -\frac{1}{2}\sum_k \frac{U(n_0 + \tilde{m})}{\hbar\omega_k} \coth\left(\frac{\hbar\omega_k}{2T}\right). \tag{7}$$

At finite temperature, these quantities are related as

$$(2\tilde{n}_k + 1)^2 - 4|\tilde{m}_k|^2 = \coth^2(\hbar\omega_k/2T). \tag{8}$$

At zero temperature, Eq.(8) reduces to

$$|\tilde{m}_k|^2 = \tilde{n}_k(\tilde{n}_k + 1). \tag{9}$$

This equation constitutes an explicit relationship between the normal and the anomalous densities at zero temperature and indicates that these two quantities are of the same order of magnitude at low temperatures which leads to the fact that neglecting $\tilde{m}$ while maintaining $\tilde{n}$ is a risky approximation.

Before calculating the normal and anomalous averages it is interesting to study their asymptotic behavior.

In the center of the Brillion zone where $k \to 0$, quantities (6) and (7) behave as

$$\tilde{n}_k = \tilde{m}_k \simeq \frac{U(n_0+\tilde{m})}{(C_s k)^2}, \tag{10}$$

where $C_s = \sqrt{2ta^2U(n_0+\tilde{m})}$ is the sound velocity.

At the boundary of the Brillion zone where $k \to \pi/a$, the spectrum takes the form

$$\hbar\omega_k \simeq 2\sqrt{\mathrm{z_0 t}[\mathrm{U}(n_0+\tilde{m}) + \mathrm{z_0 t}]}\ . \tag{11}$$

Eqs. (10) and (11) show that $\tilde{n}$ and $\tilde{m}$ are integrable and therefore the problem of divergences of the anomalous density is no longer posed in the lattice in contrast with the ordinary case where $\tilde{m}$ provides ultraviolet divergences and thus requires either normalization of the coupling constant or regularization [4,15-19]. Consequently the lattice plays the role of regularization scheme to treat ultraviolet divergences. Physically this can be justified by the fact that the atoms are localized in the Brillouin zone and thus all integrals can be calculated inside the domain $D = \left[-\frac{\pi}{a}, \frac{\pi}{a}\right]$.

In order to check whether the GHFB formalism is able to predict the quantum phase transition, it is useful to analyze the behavior of the condensed fraction as a function of $U/t$ in the limit of dilute gas where $\tilde{m}/n_0 \ll 1$ and at very low temperature where $n_0 \to n$.

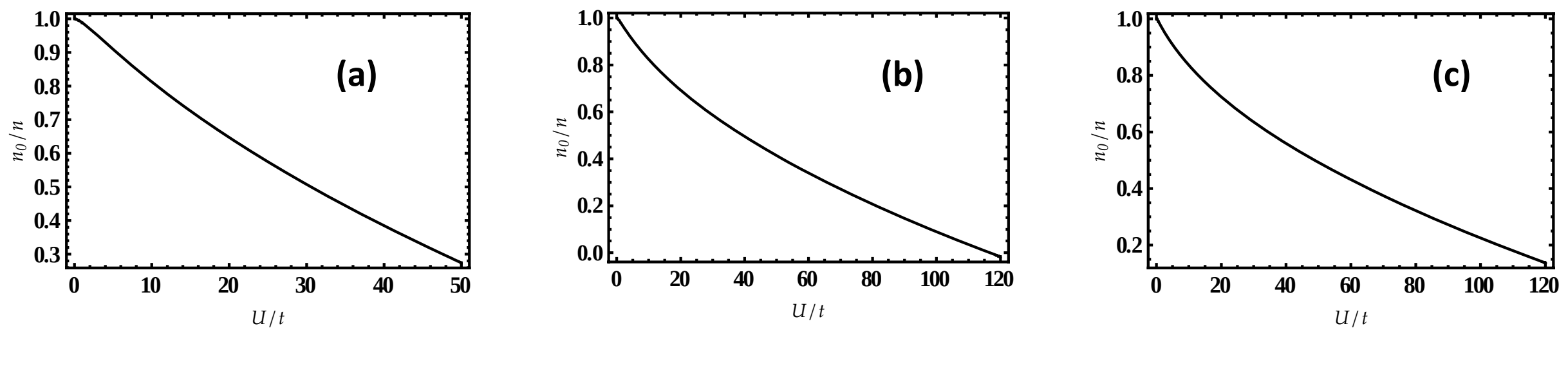

**Figure 1.Condensed fraction as function of *U/t* in three dimensional optical lattice for n=1 (a), n=2 (b) and n=3 (c).**

In figures. (1.a-c) we plot the condensed fraction $n_0/n$ versus$U/t$ for various values of $n$. We observe that for $U/t = 0$, all atoms occupy the ground state. Consequently the condensed fraction should equal to the unity. For higher values of $U/t$, $n_0/n$ decreases monotonically until it reaches zero at the critical value of $U/t$. One can observe also clearly from figures (1.b) and (1.c) that the critical value of $U/t$ increases progressively with $n$. These encouraged results lead us to deduce that the GHFB approximation has indeed succeeded in introducing a quantum phase transition. It is to be mentioned at this level that this behavior persists also in two dimensional case.

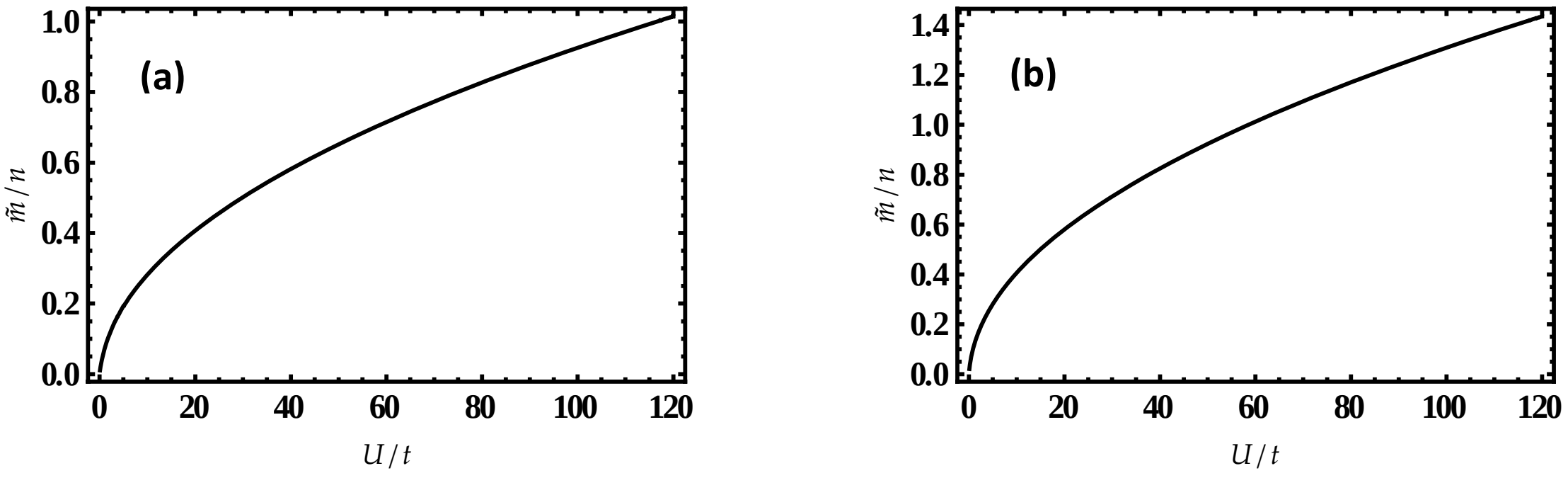

**Figure 2. Anomalous fraction as function of *U/t* in three (a) and two (b) dimensional optical lattice for n=3.**

Figure 2, depicts that the anomalous fraction increases with increasing values of $U\ /\ t$ . This is in fact natural for the anomalous density since it strictly depends to the interactions ($\tilde{m} = 0$ for $U = 0$) [19] . This feature holds also in two dimensions (see figure (2.b)) and for any value of $n$. The comparison between figures (a) and (b) shows that the central density in two dimensions is larger than the three dimensional case i.e. $\tilde{m}_{d=2}\ (0) > \tilde{m}_{d=3}\ (0)$. Thus, we can infer that the dimensionality of the system influences on the behavior of the anomalous density which is in good agreement with the ordinary 2D BEC i.e. without lattices [19].

### 3. Mott-insulator state

In this section we investigate the behavior of the noncondensed and anomalous densities in the Mott insulator phase. It is useful then to rewrite Eqs. (6) and (7) in dimensionless form. Hence, taking the continuum limit by using $\sum_k \to V \int_{-\frac{\pi}{a}}^{\frac{\pi}{a}} \frac{d^d k}{(2\pi)^d}$, we obtain after changing from momentum $k$ to a dimensionless variable $q = ak/2\pi$

$$\tilde{n} = \frac{1}{2}\int_{-\frac{1}{2}}^{\frac{1}{2}} d^d q \, \frac{\Xi_q + U(n_0+\tilde{m})}{\hbar\omega_q}\left[\coth\left(\frac{\hbar\omega_q}{2T}\right) - 1\right], \tag{12}$$

$$\tilde{m} = -\frac{U(n_0+\tilde{m})}{2} \int_{-\frac{1}{2}}^{\frac{1}{2}} \frac{d^d q}{\hbar\omega_q} \coth\left(\frac{\hbar\omega_q}{2T}\right), \tag{13}$$

where $\hbar\omega_q = \sqrt{\Xi_{\mathrm{q}}{}^2 + 2\Xi_{\mathrm{q}} U(n_0 + \tilde{m})}$.

Let us start by working at zero temperature and taking limits $U\ /\ t \to \infty$ and $\tilde{m}/n_0 \to 0$. After a little bit algebra we get for noncondensed and anomalous densities:

$$\tilde{n} = \frac{1}{4\pi}\sqrt{\frac{Un_0}{2t}}\beta_d - \frac{1}{2}, \tag{14}$$

$$\tilde{m} = -\frac{1}{4\pi}\sqrt{\frac{Un_0}{2t}}\beta_d, \tag{15}$$

with the factor $\beta_d = \int_{-1/2}^{1/2} \frac{d^d q}{|q|} = \begin{cases} 3.52549 \to d = 2 \\ 2.38008 \to d = 3 \end{cases}.$

As it is expected, one can see from Eq.(15) that the anomalous density has a large negative value owing to the strong interactions which characterize the Mott insulator state. What's important is that $\tilde{m}$ has a finite value and can be calculated easily without any divergence problem. Additionally, Eq. (14) provides a large depletion which prevents formation of the condensate in the insulating phase in both two and three dimensions. Furthermore, since the integral of Eq. (14) divergences in one dimension, the condensate cannot exist anymore in one dimensional optical lattice which well coincides with the standard Mermin-Wagner-Hohenberg theory [20, 21].

On the other hand, it has been shown recently in one of our works [19] that not only the true condensate cannot exist at finite temperature but also the anomalous density is strictly zero in a homogeneous two-dimensional Bose gas at any nonzero temperature for the reason that the long-wavelength thermal fluctuations destroy long-range order, preventing formation of both condensate and the anomalous density. It has been proven also that these two quantities arise of the symmetry-breaking assumption [4,19].

Therefore, in what follows, we confine ourselves to calculate fluctuations at nonzero temperature only in three dimensional case. It is useful then to use the expansion $\coth(x) = \frac{1}{x} + \cdots$. At high temperature when$T \to T_c$ , the normal and anomalous averages behave as [22].

$$\tilde{n}_q \simeq \frac{\Xi_k + U(n_0+\tilde{m})}{\left(\hbar\omega_q\right)^2} T - \frac{1}{2}, \tag{16}$$

$$\tilde{m}_q \simeq -\frac{U(n_0+\tilde{m})}{\left(\hbar\omega_q\right)^2}\, T. \tag{17}$$

In order to study the behavior of the noncondensed and the anomalous averages in three dimensions at any range of temperature, we solve numerically Eqs. (12) and (13) in the limit $\tilde{m}/n_0 \to 0$.

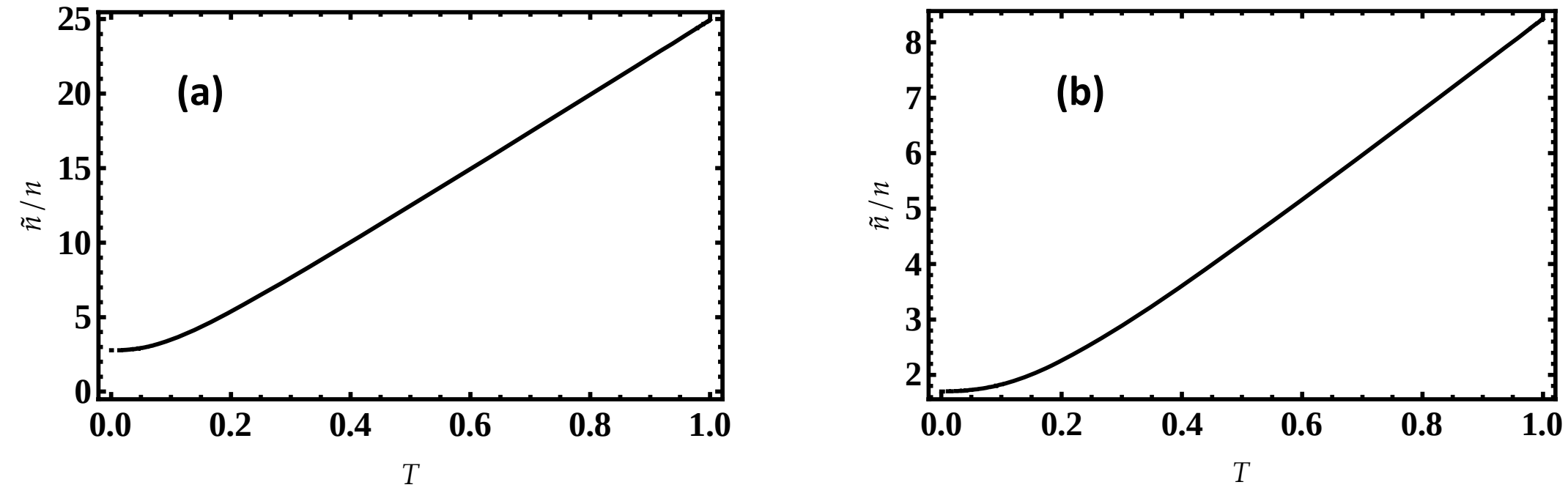

**Figure 3. Noncondensed fraction as function of $T$ in three dimensional optical lattices for $u = 1$ and $t = 0.01$ (a): $n = 1$ (b): $n = 3$.**

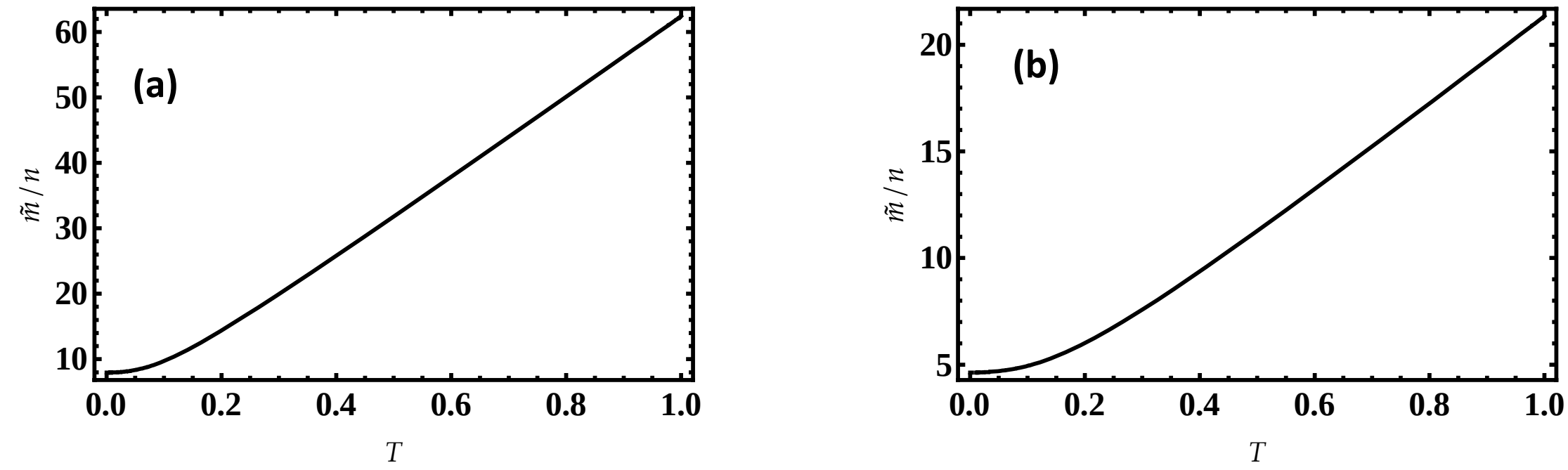

**Figure 4. Anomalous fraction as function of $T$ in three dimensional optical lattices with the same parameters as in figure.3.**

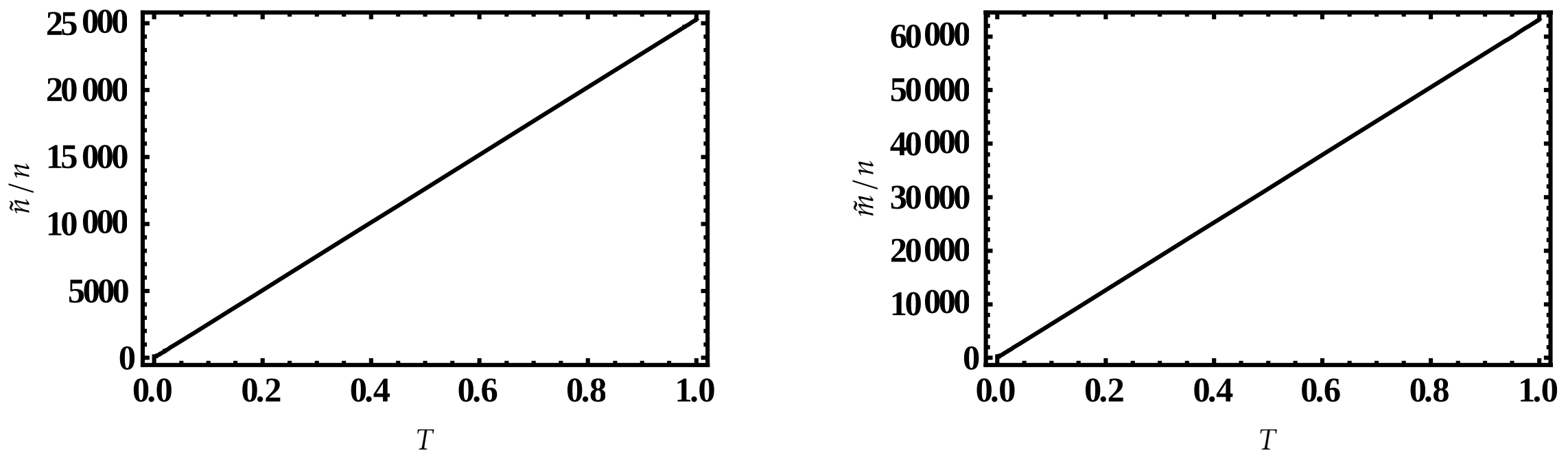

**Figure 5. Noncondensed (left panel) and anomalous (right panel) fractions as function of $T$ in three dimensional optical lattice for $n$=1, $U$=1 and $t$=$10^{-5}$.**

As can be seen from figures 3-5 that the noncondensed and anomalous fractions are increasing function versus the temperature whatever the value of $n$. For a very small value of the hopping term ($t = 10^{-5}$) i.e. $U/t \gg 1$, we observe in figure 5 that the shape of the normal and anomalous densities is enhanced. Therefore, $\tilde{n}$ and $\tilde{m}$ start to increase linearly with temperature reflecting a huge depletion and correlations in the system. Moreover, the anomalous fraction still larger than the noncondensed density at low temperature.

## 4. Conclusion:

We have studied in this paper the phase transition from the superfluid to the Mott insulator state in two and three dimensional optical lattices using the GHFB theory. The former provides a large condensed fraction with increasing values of $U/t$, which constitute encouraging results for the prediction of the phase transition. We have shown also that the anomalous density is larger than the noncondensed density in two and three dimensions at zero temperatures. Furthermore, the normal and anomalous averages increase linearly with temperature for very small value of the hoping term. We have pointed out also that the dimensionality of the lattice may enhance the behavior of the anomalous density.